\documentclass[twocolumn, twoside]{ncnsd}
\begin{document}
\title{Possibility of Prediction of Avalanches in Power Law Systems}
\author{Rumi De and G. Ananthakrishna
\thanks{Rumi De is with the Material Research Center, Indian Institute of Science,
         Bangalore-560012, India, e-mail: rumi@mrc.iisc.ernet.in,
G. Ananthakrishna is with the Material Research Center and Center
for Condensed Matter Theory, Indian Institute of Science,
Bangalore-560012, India.
         e-mail: garani@mrc.iisc.ernet.in
       }}
\markboth{NATIONAL CONFERENCE ON NONLINEAR SYSTEMS \& DYNAMICS}
{INDIAN INSTITUTE OF TECHNOLOGY, KHARAGPUR
721302, DECEMBER 28-30, 2003}
\maketitle

\begin{abstract}
We consider a modified Burridge-Knopoff model with a view to
understand results of acoustic emission (AE) relevant to
earthquakes by adding a dissipative term which mimics bursts of
acoustic signals.  Interestingly, we find   a precursor effect in
the cumulative energy dissipated which allows identification of a
large slip event. Further, the AE activity for several large slip
events follows a universal stretched exponential behavior with
corrections in terms of time-to-failure. We find that many
features of the statistics of AE signals such as their amplitudes,
durations and the intervals between successive AE bursts obey
power laws consistent with recent experimental results. Large
magnitude events have different power law from that of the small
ones, the latter being sensitive to the pulling speed.
\end{abstract}


Causes of failure of materials  and the possibility of predicting
them  is of interest in science and engineering (electrical
breakdown, fracture of laboratory samples to engineering
structures, etc). This is particularly important in seismology due
to the enormous damage earthquakes can cause. Indeed, predicting
earthquakes has been of interest to geophysics for a long time. At
a practical level, this amounts to identifying useful precursors
at a statistically significant level.  The absence of useful
precursors could possibly arise due to the inherent limitations
set by measurement processes \cite{Main96}. Even so, there has
been records of individual earthquakes where precursor effects
have been reported \cite{Main96}. Some insight into the dynamics
of earthquakes has been possible by mapping the problem to
fracture processes. In the case of fracture, the  nucleation and
propagation of cracks culminates in the failure of the material.
In such situations, the nondestructive nature of acoustic emission
(AE) is a very convenient tool for studying the process as the
emitted signals are sensitive to  the microstructural changes
taking place inside the system \cite{Mogi62,Scholz68a}. Such
studies have shown that it is possible to follow the nucleation
and growth of fracture by imaging the fracture process (through an
inversion process of arrival times) \cite{Lockner91}. In addition,
the statistics of the AE signals exhibit a power law
\cite{Mogi62,Lockner91,Scholz68b} similar to the
Gutenberg-Richters law for the magnitudes of earthquakes. 
One aim of these studies has been to look for
precursor effects \cite{Main96,Sam92}. Thus, most laboratory
studies on AE relevant to earthquake dynamics are on rock samples
subjected slip with appropriate geometry \cite{Scholz68a}.

Apart from the power laws observed in AE signals during fracture,
acoustic activity of unusually large number of situations exhibit
power laws in systems as varied as volcanic activity
\cite{Diodati}, microfracturing process \cite{Petri94,Garci97},
collective dislocation motion \cite{Weiss} and martensitic
transformation \cite{Vives} to name a few.  Though the general
mechanism attributed to AE is the release of  stored strain
energy, the details are system specific. Thus, the ubiquity of the
power law statistics of AE signals suggests that the details of
the underlying processes are irrelevant. One framework which
unifies such varied situations is   that of the self-organized
criticality (SOC) \cite{Bak}. This approach has been  successful
in explaining the statistical self similarity of the seismic
process reflected in the Gutenberg-Richter's law for earthquake
magnitudes \cite{Bak89}, as also the power laws in other systems
\cite{Diodati,Petri94,Garci97,Weiss,Vives,Rajeev}. However, given
that earth is a SOC state,  at a conceptual level, doubts have
been raised about the possibility of predicting an earthquake.  A
recent debate on this subject concluded that deterministic
prediction of individual earthquakes is unrealistic
\cite{Debate99}.  Clearly, the lack of predictability is
applicable to all power law systems. However, in the general
context of failure, time-to-failure models have been used for the
prediction of failure \cite{Huang98}. These models are thought to
be applicable to earthquake like situations also. There are some
theoretical efforts to look for precursor effects before the onset
of large avalanches in SOC type models as well
\cite{Rosen94,Bikas96}.

The power law  statistics of the AE  signals (on rock samples) is
believed to result from  slip events. However, to the best of
knowledge there are no phenomenological models that mimic AE
bursts. This also helps us to identify a possible precursory
effect. One other interesting feature of earthquake magnitudes is
the change in the power law exponent for small and large
magnitudes \cite{Scholz92,Sornette}. Laboratory studies on AE
signals on rock samples also appear to indicate such a change
\cite{Yabe}. Finally, recent studies on AE also show that the
exponent value is sensitive to the deformation rate
\cite{Yabegrl}. To the best of our knowledge, there has been no
explanation of these observations. Here, we introduce an
additional dissipative term into the Burridge-Knopoff (BK) model
which captures the main features of AE signals.

It is known that  slip events result due to deformation and /or
breaking of asperities  resulting in an accelerated motion of the
local areas of slip. Since AE is due to the release of the
built-in strain energy as the system  surmounts the threshold, we
consider this accelerated motion during a local slip as
responsible for acoustic emission. Such a rapid movement prevents
the system from attaining a quasi-static equilibrium. This also
implies that there are dissipative forces that resist this abrupt
motion. We introduce the Rayleigh dissipation functional which
depends on the gradient of the local strain rate \cite{Land} to
account for the dissipation arising from the rapidly moving blocks
in the BK model for earthquakes \cite{BK,Carlson}. Indeed, such a
dissipative term has been successful in explaining the power law
statistics of the AE signals during martensitic transformation
\cite{Rajeev}.

The Burridge-Knopoff model for earthquakes \cite{BK} and its
variants have been studied in detail by many authors
\cite{Carlson}. It consists of a chain of blocks of mass $m$
coupled to each other by coil springs of strength $k_c$ and
attached to a fixed surface by leaf springs of strength $ k_p$
(Fig. 1a). The blocks are in contact with a rough surface moving
at constant speed $V$. The velocity-dependent friction force `$f$'
operates between the blocks and the surface. We use two forms of
frictional force schematically shown by the solid and dashed
curves shown in Fig. 1b.

\begin{figure}[t]
\hbox{
\includegraphics[height=2.0cm,width=4.5cm]{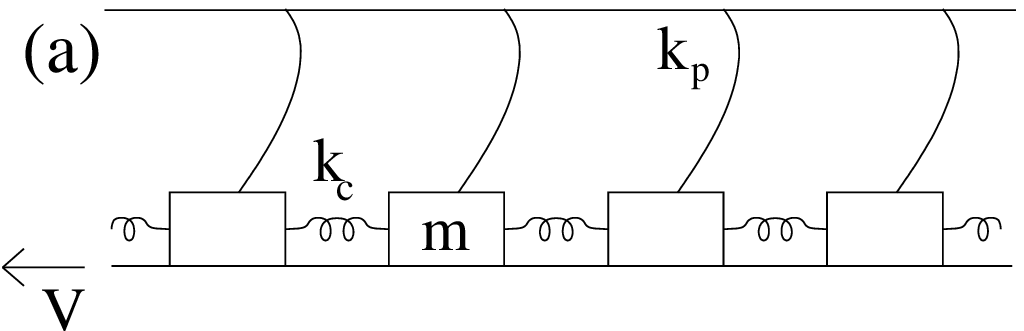}
\hspace{0.5cm}
\includegraphics[height=2.2cm,width=3.5cm]{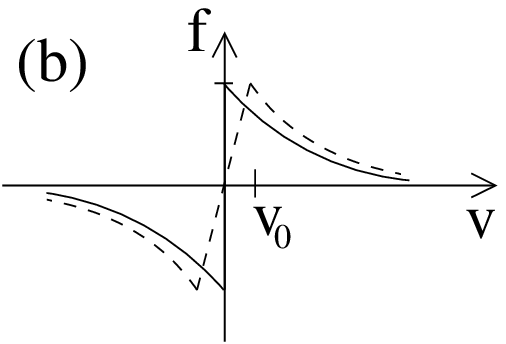}}
\caption{(a) The Burridge-Knopoff spring block model. (b)Friction laws: The coulomb friction law at zero velocity beyond which a smooth velocity weakening friction law (solid line). Dashed curve describes a creep branch with a similar velocity-weakening behavior. }
\label{}
\end{figure}

Here, we introduce an dissipation associated with the rapid slip
events represented by the Rayleigh dissipative functional
\cite{Land} $R = {\gamma_c\over{2}}\int({{\partial\dot
u(x)}\over{\partial x}})^2dx$. Then, in the notation of Ref
\cite{Carlson}, the equations of motion can be written as

\begin {eqnarray}
\ddot U_j &=& l^2(U_{j+1} - 2U_j + U_{j-1} ) - U_j - \phi (2 \alpha
\nu + 2\alpha \dot U_j) \nonumber \\
&+& {\gamma_c} ( \dot U_{j+1} - 2\dot U_j +
\dot U_{j-1} ),
\end {eqnarray}
\noindent where $ U_j$ is the dimensionless displacement of the
$j^{th}$ block, $\nu $ is the dimensionless pulling velocity, the
ratio of the slipping time to the loading time, $l^2 $=$ k_c/k_p$
represent the stiffness ratio and the parameter describing the
rate of velocity-weakening in the friction is $\alpha$,
The last term arises from $R(t)$, the additional dissipative term
introduced to mimic the AE bursts and $\gamma_c$  is the scaled
dissipation coefficient. ( The over dot refers to differentiation
with respect to dimensionless time  variable.)

This model without the last term has been extensively studied
\cite{BK,Carlson}. Starting from random initial conditions, slip
events ranging from one-block event to those extending over the
entire fault (`large events', occurring roughly over one loading
period of $\tau_L \sim 2/\nu$ ) are seen in the steady state.
These earthquake-like events mimic the empirical Gutenberg-Richter
law.

Simulations have been carried out using fourth-order Runge-Kutta
method with open boundary condition.  Random initial conditions
are imposed. After discarding the initial transients, long data
sets are recorded when the system has reached a stationary state.
The parameters used here are $l=10, \alpha = 2.5, N = 100$ for two
sets of values of $\nu = 0.01$ and 0.001 for range of values of
$\gamma_c$. The calculations have been carried out for both the
frictional laws shown in Fig. 1b. In the creep  case, the creep
branch ends at a value of $\nu \sim 10^{-7}$ beyond which the
velocity weakening law operates. The modified BK model produces
the same statistics of  slip events as that without this term as
long as the value of $\gamma_c$ is small, typically
$\gamma_c < 0.5$. The results presented here are for $\gamma_c =
0.02$.

\begin{figure}[htb]
\vbox{
\includegraphics[height=3.5cm,width=8.5cm]{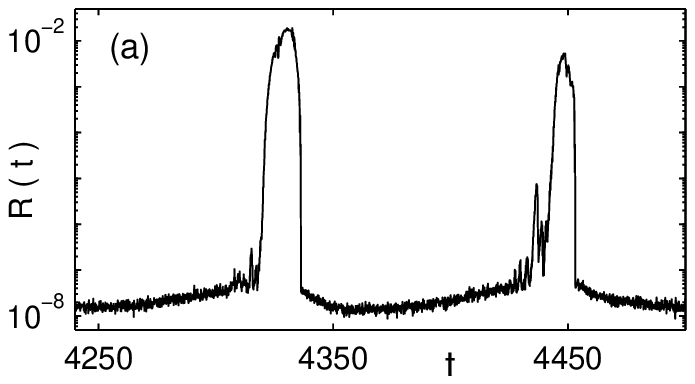}
\includegraphics[height=4.3cm,width=8.5cm]{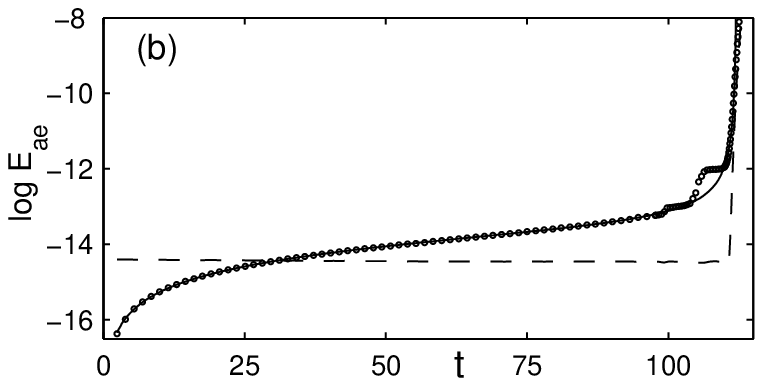}}
\caption{(a) R(t) as a function of t.  (b) $logE_{ae}$ as a
function of time t ($\circ$) along with the fit (solid line) for
the friction law with creep branch. Dashed line corresponds to the mean
kinetic energy.}
\label{}
\end{figure}
Since the rate of energy dissipated \cite{Land} due to local
accelerating blocks is given by $dE_{ae}/dt = -2R(t)$, we
calculate $R(t) =(\gamma_c/2)\sum_{j} (\dot U_j - \dot
U_{j-1})^2$. We find that the energy dissipated occurs in bursts
which is similar to the acoustic emission signals. A plot of
$R(t)$ is shown in the Fig. 2a for the case when the frictional law
has a creep branch. {\it A gradual increase in the activity of the
energy dissipated can be seen to accelerate just prior to the
occurrence of a 'large'  slip event.} This feature is seen for all slip events
of observable magnitude which we refer as large.
This suggests that $R(t)$ can be used as a precursor for the onset of a
slip event observed in experiments on rock samples \cite{Sam92}.
As the energy dissipated is noisy, a better quantity for the
analysis is the cumulative energy dissipated $E_{ae}(t)$ (
$\propto \int_0^t R( t^{\prime} ) d t^{\prime} $). This grows in a
stepped manner and as we approach a slip event, $ E_{ae}$,
increases rapidly with the steps becoming increasingly visible. A
plot of $log E_{ae}$ is shown in Fig. 2b along with a fit having
the functional form (continuous line),
\begin{equation}
log E_{ae}(t) =  -a_1 t^{-\alpha_1} [ 1 - a_2 \vert (t-t_c)/t_c
\vert ^{-\alpha_2}].
\end{equation}
Here, the crucial parameter $t_c$ is the time of occurrence of the
event and $t$ is time measured from some initial point after a
slip event. The constants $a_1,a_2,\alpha_1$ and $\alpha_2$ are
determined by a  fit to the data. It is clear that the fit is
striking. Given a reasonable stretch of the data, the initial
increasing trend in $log E_{ae}$ is easily fitted to a stretched
exponential, ie., $- a_1 t^{-\alpha_1}$. The second term is
introduced to account for the observed rapid increase in the
activity as we approach the time of failure. As the mean kinetic
energy is a good indicator of the failure time, we have shown this
by a dashed line.  It is clear that the estimated $t_c$  agrees
quite well with that of the mean kinetic energy.

\begin{figure}[t]
\includegraphics[height=4.3cm,width=8.5cm]{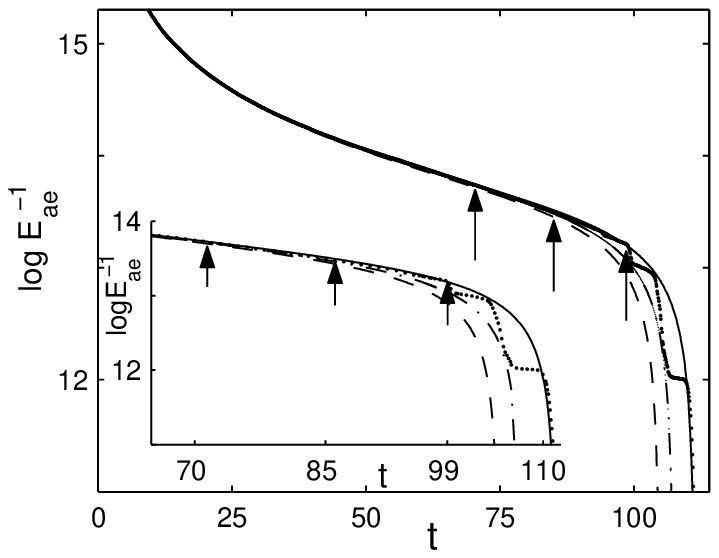}
\caption{ A plot of $logE_{ae}^{-1}$ versus t. Inset shows the enlarged section at
 time $t_1$ ($-$), $t_2$ ($-\cdot$) and $t_3$ (solid line). Data shown ($\cdot$) is indistinguishable from the fit till $t_3$.}
\label{invR}
\end{figure}

\begin{figure}{t}
\includegraphics[height=4.0cm,width=8.5cm]{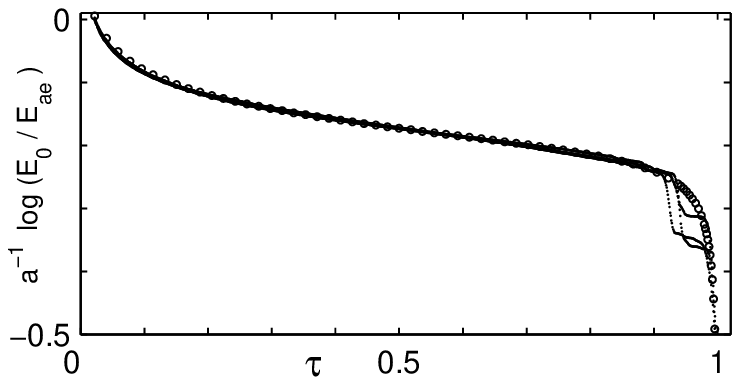}
\caption{Collapsed data using $ a_1^{-1} log (E(0)/E(\tau))$ versus $\tau$ for three different events along with the fit shown by ($\circ$) for friction law with creep.}
\label{univ}
\end{figure}

A proper estimate of the  warning time requires that the values of
the four constants $a_1,a_2,\alpha_1$ and $\alpha_2$ do not change
in time significantly. Indeed, we find that these constants change
very little given a data $E_{ae}(t)$ over a reasonable initial
stretch of time. Only $t_c$ changes. Still, we are left with the
problem of obtaining a best estimate of $t_c$. This is done as
follows. Consider the plot of $E_{ae}^{-1}(t)$ shown in Fig.
\ref{invR}. Given the data till $t= t_1$ ( the first arrow shown
in Fig. \ref{invR} ), we find that the four parameters  are
already well determined (within a small error bar). A fit to Eq.
(2) also gives $t_c^{(1)}$  at $t_1$.  One such curve is shown by
a dashed line with the arrow shown at $t_1$. The value of
$t_c^{(1)}$ is only an estimate based on the data till $t_1$.
However, as time progresses, the data accumulated later usually
deviates from the predicted curve beyond $t_1$ if $t_c^{(1)}$ is
inaccurate. This is case for the fit till $t_1$ and $t_2$  for
instance  shown in Fig. \ref{invR}. If on the other hand, the
deviation of the predicted curve from the accumulated data
decreases with passage of time within the error bar, then, the
value of $t_c$ is likely to be accurate. Indeed, the extrapolated
(continuous) curve shown in Fig. \ref{invR} corresponding to data
fit till $t=t_3$ with the corresponding predicted $t_c^{(3)}$ is
seen to follow the data very well. (Usually, this is followed by a
sudden decrease in $E_{ae}^{-1}$ which is again an indication of a
possible large event, but the general trend soon follows the
extrapolated curve.) Then, one can take $t_3$ to be the warning
time of the impending large event. From the data shown in Fig.
\ref{invR}, the actual $t_c$ is 110.4 where as the predicted $t_c$
is 111.6. Thus, the percentage error in the prediction of the
onset of the a large event $\sim 1\%$. The power law nature of
$E_{ae}$ also suggests that the approach to different events is
universal. We find that the data corresponding to different events
collapses into a single curve given by $ a_1^{-1} log
[E_{ae}(0)/E_{ae}(\tau)] = \tau^{-\alpha_1} [1 -a_2 \vert (\tau
-1)^{-\alpha_2}]+a_{1}^{-1}logE_{ae}(0) $ in terms of $\tau =t/t_c$. This is shown in
Fig. \ref{univ} along with the fit for three different events.

\begin{figure}
\includegraphics[height=4.3cm,width=8.5cm]{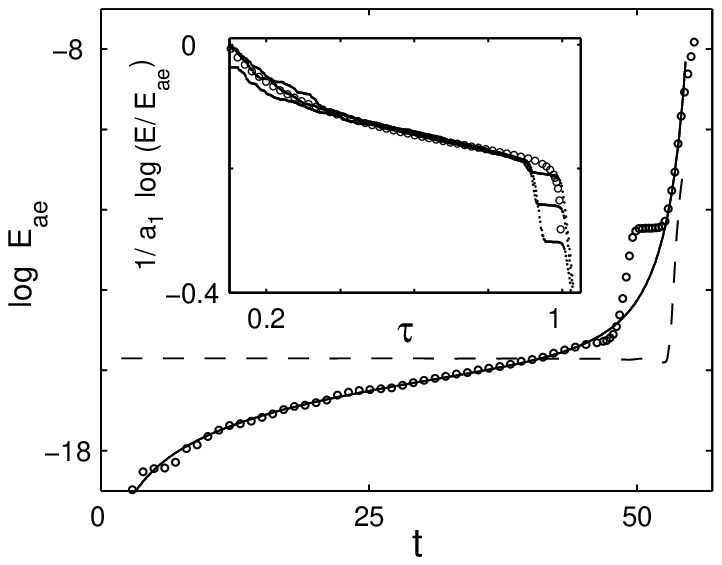}
\caption{ $logE_{ae}$ as a function of time  t ($\circ$) along with a fit (solid line) for the coulomb friction . Dashed line indicates the mean kinetic energy. The inset shows the collapsed curve for three different events along with a fit ($\circ$).}
\label{coul}
\end{figure}

For the Coulomb friction law, we find that $R(t)$ is much more
noisy. A plot of $log E_{ae}$ for one large event is shown in Fig. \ref{coul}. Following the same procedure, we find that  we can fix only
$a_1$ and $\alpha_1$ reasonably well given an initial stretch of
data. However, the parameters $a_2$ and $\alpha_2$ also converge
within some error bar which is more than the creep case. Although,
the changes in $t_c$ with $t$ is more than the previous case, the
$t_c$ is fixed the same way, but the error is larger than the
previous case. The data along with the fit (up to the point shown
by arrow) is clearly seen to mimic the rapid increase in $log
E_{ae}$. The data collapse for several such curves (for $ a_1^{-1}
log [E_{ae}(0)/E_{ae}(\tau)]$ ) is also shown in the inset along
with the fit. (The scatter in the collapsed data is reasonable
except for the initial part of the data.) However, we do find that
this procedure does pose problems in a few specific set of events
for which approach to the large event is very noisy.

\begin{figure}[tbh]
\vbox{
\includegraphics[height=4.3cm,width=8.5cm]{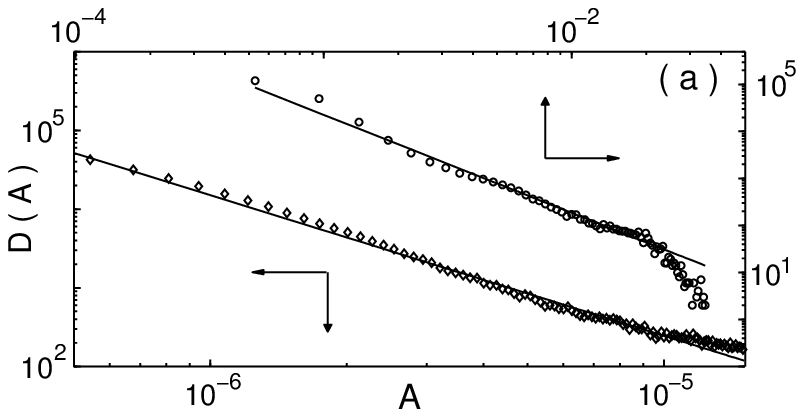}
\includegraphics[height=4.0cm,width=8.3cm]{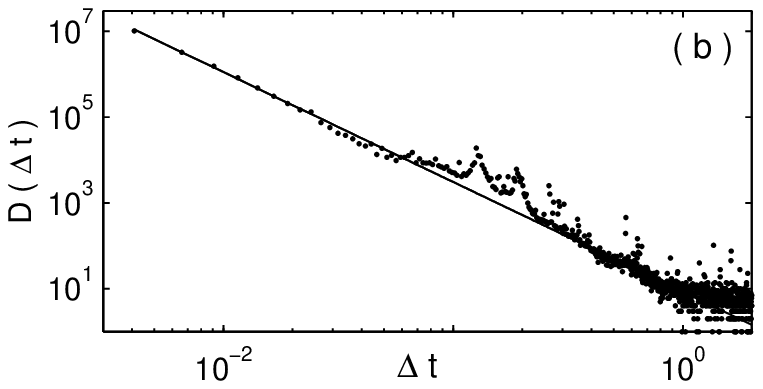}}
\includegraphics[height=4.3cm,width=8.5cm]{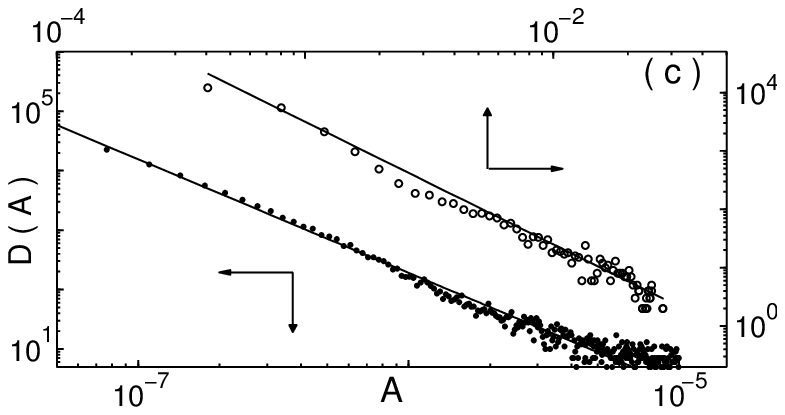}
\caption{(a)Distribution of amplitude of $R(A)$, $D(A)$ versus A.
($\diamondsuit$) indicates small amplitudes and (o) shows large
amplitudes for $\nu$ = 0.01, (b) Distribution of time intervals of
AE events and (c) $D(A)$ versus A for $\nu$ = 0.001, ($\bullet$) corresponds to the small amplitudes and (o) that for large.} \label{powerlaw}
\end{figure}

We now consider the statistics of the energy bursts. The
distribution of the magnitudes of $R(t)$ shows a power law
$D(A)\sim A^{-m}$, where  $D(A)$ is the number of events between
$A$ and $A+dA$. Instead of a single power law  anticipated, we
find that the distribution shows two  regions, one for relatively
smaller amplitudes and another for large values shown by two
distinct plots in Fig. 6a. ( This is for the Coulomb frictional
law for $\nu = 0.01$. Similar results are obtained when the
frictional law has a creep branch.) The value of $m$ for the small
amplitudes region ( $ < 10^{-4}$) is $\sim 1.78$, significantly
smaller than that of large amplitudes which is $\sim 2.09$. This
qualitative feature is consistent with the recent experimental
results on rock samples \cite{Yabe}. Indeed, this is similar to
the  well known observation in the case of seismic moments where a
deviation from the power law for lower magnitudes beyond a certain
value ( $> 7.0$ on the Richter scale \cite{Scholz92,Sornette}) is
noted. Finally, one other quantity of interest is the time
interval between the events which we have calculated. This is
shown in Fig. 6b for $\nu =0.01$. There appears to single power
law with and exponent 2.6. However, there is considerable scatter
in the mid region corresponding crossover in the power laws from
small to large amplitude region. For smaller pulling speeds, we
see two different scaling regimes as for the amplitude.

Recently  Yabe et. al. \cite{Yabegrl} have noted  that the
exponent value in the small amplitude region increases with
decreasing deformation rate in contrast to the large amplitude
region. In order to check this, we have performed run for $\nu
=0.001$ for which the data (for the Coulomb friction law ) is
shown in Fig. 6c. While the exponent for small amplitude regimes
is 1.91 higher than that for $\nu = 0.01$, the exponent for larger
amplitudes is insensitive. This result can be physically explained
by analyzing the influence of the pulling velocity on slip events
of varying sizes. We first note that Rayleigh dissipation function
is the gradient of local slipping rates. From the arguments
presented in Ref. \cite{Carlson}, one knows that the velocity of
one block event is proportional to $\nu$. Further, as the
neighboring blocks are at rest, the number of such events are
fewer in proportion to the pulling speed, both of which are
evident from Fig. 6a and Fig. 6c. When we consider the two block events, it
clear that the difference in the velocities of the two blocks
being of similar magnitude contributes very little and only edges
contribute. In a similar way, it can be argued that for slip
events of finite size, the extent of the contribution to $R(t)$ is
decided by the ruggedness of the velocity profile within the
slipping region; it is lower if the velocity is smoother. The
ruggedness of the velocity profile, however, is itself decided by
how much time the system gets to 'relax'. At lower pulling
velocities, there is sufficient time for the blocks to relax
compared to higher pulling  velocities.  Thus, the velocity
profile within slip event of certain magnitude tends to be much
more smooth at low $\nu$ compared to at higher higher $\nu$ as in
the latter case it does not allow for full relaxation (to attain
near quasi-stationary state).

In summary, within the scope of this model, we have shown that
acoustic emission could be used as a possible precursor for
detecting an event during the process of failure. In the case
where the friction includes a creep branch, we find that the time
of failure can be predicted quite accurately. In the coulomb case,
the energy dissipated is quite noisy. Even then, the predicted
failure time is quite reasonable. At the first sight, the
predictability aspect appears to be surprising considering the
fact the statistics of the events exhibits a power law. However,
the data collapse for different events clearly suggests that the
dynamics of approach to large events is itself universal and scale
invariant. This scale invariant form implies that all events of
detectable magnitude is describable by the same equation, the only
limitation being the ability to detect. One limitation of Eq. (2)
is that the magnitude of the energy dissipated appears to bear no
correlation with the magnitude of the slip event as larger events
often show higher $R(t)$.  For instance, even when the ratio of
the kinetic energies between two events is four orders smaller,
the energy dissipated $R(t)$ in the two cases do not scale in
proportion. For the same reason, the model is unable to predict
the magnitude of the event. We stress that this precursor effect
is absent in the total kinetic energy or other variables. We also
expect these results to be applicable at the laboratory level in
failure of materials and structures. It is worth noting that the
form of approach to failure is different from that given by Huang
et al \cite{Huang98}. Apart from this, the model also predicts
that there are two exponent values one for small amplitudes and
another for large amplitudes consistent with experimental results
\cite{Yabe}. We also find that the exponent value corresponding to
low amplitudes to be much more sensitive to the pulling speed than
that at large amplitudes. This dependence on the pulling speed has
been  be traced to the form of $R(t)$, namely, the gradient of the
local strain rate. Finally, our analysis shows that the exponent
values of the bursts of acoustic energy is not the same as that of
the event size distribution (as used in Ref.\cite{Carlson}).

The authors wish to acknowledge the support from department of science and technology.

\bibliographystyle{ncnsd}

\end{document}